\documentclass{amsart}

\usepackage{amssymb,amsthm,amsmath,graphicx}

\newtheorem{thm}{Theorem}[section]
\newtheorem{prop}[thm]{Proposition}
\newtheorem{lem}[thm]{Lemma}

\theoremstyle{definition}
\newtheorem{definition}[thm]{Definition}

\newtheorem{assumption}{Assumption}
\newtheorem{remark}{Remark}

\numberwithin{equation}{section}

\newcommand{\norm}[1]{\left\Vert#1\right\Vert}
\newcommand{\abs}[1]{\left\vert#1\right\vert}
\newcommand{\set}[1]{\left\{#1\right\}}
\newcommand{\eps}{\varepsilon}
\newcommand{\real}{\mathbb{R}}  
\newcommand{\integ}{\mathbb{Z}}

\title{Stability of the inverse resonance problem for Jacobi operators}
\author{Matthew Bledsoe}
\address{Department of Mathematics, University of Alabama at Birmingham, Birmingham, AL 35294, USA}
\email{bledsoem@uab.edu}

\begin{document}
\begin{abstract}
When the coefficients of a Jacobi operator are finitely supported perturbations of the $1$ and $0$ sequences, respectively, the left reflection coefficient is a rational function whose poles inside, respectively outside, the unit disk correspond to eigenvalues and resonances.  By including the zeros of the reflection coefficient, we have a set of data that determines the Jacobi coefficients up to a translation as long as there is at most one half-bound state.  We prove that the coefficients of two Jacobi operators are pointwise close assuming that the zeros and poles of their left reflection coefficients are $\eps$-close in some disk centered at the origin.
\end{abstract}

\maketitle

\section{Introduction}\label{sec:intro}
The inverse resonance problem for a variety of operators has garnered interest in the past decade \cite{Zworski2001,Brown2003,Brown2004,Korotyaev2004,Rundell2004,Brown2005,
Korotyaev2005,Brown2009,Weikard2010,Korotyaev2011,AW11}.  This problem concerns recovering the coefficients of a differential or difference operator (e.g. the potential of a Schr\"odinger operator) from a set containing the eigenvalues and resonances.  The concept of resonances arises from physics where they are understood as \emph{metastable states} \cite{Zworski1999}.  Mathematically, they are defined as the poles in the so-called unphysical sheet of the meromorphically continued scattering matrix.  The poles in the physical sheet are the eigenvalues of the operator.  The initial interest in using eigenvalues and resonances as data for the inverse problem, as opposed to the standard inverse scattering data, is that these objects (at least in the Schr\"odinger case) can (in principle) be measured in a lab.

Stability is a fundamental question in inverse problems.  However, compared with uniqueness, it has received little attention in the context of inverse spectral/scattering problems.  Stability results for the inverse resonance problem can be found in \cite{Korotyaev2004a,Marletta2007,Marletta2009,MNSW10,Bledsoe2012}.  For the sake of completeness, we also mention stability results for inverse spectral/scattering problems \cite{Ryabushko1983,Aktosun1987,McLaughlin1988,Dorren1994,Marletta2005,Hitrik2000,Horvath2010}.  

The present work concerns the inverse resonance problem for Jacobi operators defined on $\integ$.  Specifically, we consider the equation
\begin{equation*}
a_{n-1}f(n-1,z)+a_nf(n+1,z)+b_nf(n,z)=(z+z^{-1})f(n,z)
\end{equation*}
for $n\in\integ$ and $z\in\mathbb C$.  The classical inverse scattering problem requires the left (or right) reflection coefficient, the eigenvalues, and norming constants to uniquely reconstruct the coefficients $a$ and $b$, see \cite{Teschl2000}.  However, if $a-1$ and $b$ are finitely supported,  then by Lemma~\ref{lem:refcoeffonly} the (left) reflection coefficient is sufficient for uniqueness.  In this case, the reflection coefficient is a rational function of $z$ whose poles inside (resp. outside) the unit disk correspond to eigenvalues (resp. resonances).  The question then becomes: do the eigenvalues and resonances determine the reflection coefficient?  The answer is, of course, no, since we need (at least) the \emph{zeros} of the reflection coefficient, as well.  It is this set, the zeros and poles of the reflection coefficient, which we will use as data.

Using the zeros and poles of the reflection coefficient, we find (\cite{Korotyaev2011}, Theorem~\ref{thm:unique}) that the sequences $a$ and $b$ are determined up to a translation except in an exceptional case which is explained in Section~\ref{sec:scat}.  By fixing the minimum of the supports of the coefficients and avoiding the exceptional case, our result is as follows.  Supposing two Jacobi operators have reflections coefficients whose zeros and poles are $\eps$-close in the disk $\abs{z}<R$, we  obtain a pointwise bound of the difference of the coefficients of the two operators which depends on $\eps$ and $R$.  This result may be viewed as a discrete analogue of the author's work \cite{Bledsoe2012} as the ideas for the proof are similar.  We also employ methods based upon those appearing in \cite{Marletta2007,MNSW10} where the case of Jacobi operators defined on $\mathbb N$ is treated.

In other contexts \cite{Marletta2007,MNSW10,Bledsoe2012}, this type of result is called a ``finite data'' stability result, because the reflection coefficient in these cases has infinitely many poles or zeros.\footnote{In the half-line case considered by Marletta \emph{et al.}, the zeros of the Jost function, the eigenvalues and resonances, are the data.}  However, in our case, there are only finitely many zeros and poles, so an explanation of why we frame our result the way we do is necessary.  

Consider the following example.  Let $\delta_n$ be the sequence which is 1 at $n$ and 0 otherwise, and let $b=\delta_0$ and $\tilde b=\delta_0+0.0001\delta_5$.  In addition, let $a$ and $\tilde a$ both be the $1$ sequence.  Then, $a$ and $b$ and $\tilde a$ and $\tilde b$ define two Jacobi operators whose coefficients are pointwise close.  Their respective sets of eigenvalues and resonances (i.e., poles of their reflection coefficients) are
\begin{align*}
&\{-1.618,0.618\}\\
&\{-2.875,-1.627,0.618, 3.198, 2.296\pm 2.292i,0.1059\pm3.25i,-2.059\pm 2.337i\}.
\end{align*}
Note that the two elements of the first set are very close to the second and third elements of the second set, while the remaining elements of the second set all have modulus greater than 2.8.  The extra resonances result from the difference between the maximum and minimum of the support of $\tilde b$.  From this example, we see that requiring \emph{all} the zeros and poles to be $\eps$-close would be too restrictive.

The outline of the paper is as follows.  In Section~\ref{sec:scat} we set notation, review scattering theory for Jacobi operators, and prove the uniqueness result, Theorem \ref{thm:unique}, upon which the the stability analysis in Section \ref{sec:stability} is based.  In Section~\ref{sec:main}, we prove the main result, Theorem~\ref{thm:stability}, after establishing the necessary inequalities in sections \ref{sec:transop} and \ref{sec:wands}.

\section{Scattering theory}\label{sec:scat}
We begin this section with some well--known facts about scattering theory for Jacobi operators; a standard reference is \cite[Chap. 10]{Teschl2000}.  The different spaces of sequences will be written as follows: the set of all complex-valued sequences is $\ell(\integ)$; for $p\in[1,\infty]$, the usual Banach (Hilbert for $p=2$) spaces are $\ell^p(\integ)$; and the set of all finitely supported sequences is $\ell_0(\integ)$.  Each of the previous sets has a corresponding subset of real-valued sequences written $\ell(\integ,\real)$, $\ell^p(\integ,\real)$, and $\ell_0(\integ,\real)$.  

Let $a,b\in\ell(\integ,\real)$ such that $a_n\neq 0$ for all $n\in\integ$.  The Jacobi difference expression (JDE) $\tau:\ell(\integ)\to\ell(\integ)$ defined by $a$ and $b$ is
\begin{equation*}
(\tau f)(n)=a_{n-1}f(n-1)+a_nf(n+1)+b_nf(n).
\end{equation*}
We reserve the notation $J$ for the operator $J:\mathcal D(J)\to\ell^2(\integ)$ with domain $\mathcal D(J)=\set{f\in\ell^2(\integ):\tau f\in\ell^2(\integ)}$ and defined by $Jf=\tau f$.  For $a\equiv 1$ and $b\equiv 0$, the free Jacobi difference expression and operator are labelled $\tau_0$ and $J_0$, respectively.

Though many of the results in this section hold under more general conditions (see \cite{Teschl2000}), we make the following assumption, in force throughout the rest of the paper unless explicitly noted.  

\begin{assumption}\label{assume:finitesupp}
The sequences $a$ and $b$ defining $\tau$ satisfy: 
\begin{itemize}
\item[(i)]$a-1,b\in\ell_0(\integ,\real)$;
\item[(ii)] $a_n>0$ for all $n\in\integ$;
\item[(iii)] there are integers $N^\pm$ such that supp $b\subset\set{N^-,\dots,N^+}$ and supp $(a-1)\subset\set{N^-,\dots,N^+-1}$.
\end{itemize}
\end{assumption}

Let $\tau$ satisfy Assumption~\ref{assume:finitesupp}.  Then, for every $z\in\mathbb C\setminus\set{0}$ there exist unique solutions, $f^\pm$, of the equation
\begin{equation}\label{dse}
(\tau f^\pm)(n,z)=(z+z^{-1})f^\pm(n,z)
\end{equation}
such that
\begin{equation}\label{jostbc}
\begin{aligned}
f^+(n,z)&=z^n,\quad n\ge N^+\\
f^-(n,z)&=z^{-n},\quad n\le N^-.
\end{aligned}
\end{equation}
These solutions are called the \emph{Jost solutions} for $\tau$.  Define 
\begin{equation}\label{A}
A^+(n)=\prod_{m=n}^\infty a_n,\quad A^-(n)=\prod_{m=-\infty}^{n-1} a_n,\quad A=A^+(n)A^-(n)=\prod_{m=N^-}^{N^+-1}a_n.
\end{equation}
Note that $A^\pm(n)=1$ for $\pm(n-N^\pm)\ge 0$.  

\begin{lem}\label{lem:poly}
For each $n\in\integ$, the functions $g^\pm(n,\cdot)$ defined by
\begin{equation*}
g^\pm(n,z)= A^\pm(n)z^{\mp n}f^\pm(n,z)
\end{equation*}
are polynomials with real coefficients such that $g^\pm(n,0)=1$.  The degree of $g^+(n,\cdot)$ is at most $2(N^+-n)-1$ for $n<N^+$, and the degree of $g^-(n,\cdot)$ is at most $2(n-N^-)-1$ for $n>N^-$.  
For $\pm(n-N^\pm)\ge 0$, we have $g^\pm(n,z)=1$ for all $z$.
\begin{proof}
The final claim is an immediate consequence of \eqref{jostbc} and \eqref{A}.  We prove only the remaining claims for $g^+$ since the ones for $g^-$ are similar.  

Solving \eqref{dse} for $f^+(n-1,z)$ and multiplying through by $A^+(n-1)z^{-(n-1)}$ yields
\begin{equation*}
g^+(n-1,z)=g^+(n,z)(1+z^2-b_nz)-a_n^2z^2g^+(n+1,z).
\end{equation*}
Thus, we see that 
\begin{equation*}
g^+(N^{+}-1,z)=1-b_{N^+}z
\end{equation*}
which has the desired properties.  Proceeding by strong induction on $N^+-n$ completes the proof.
\end{proof}
\end{lem}

By the preceding lemma, there are numbers $K^\pm_0(n,n\pm j)\in\real$ such that
\begin{equation}\label{g}
g^\pm(n,z)=1+\sum_{j=1}^{\pm2(N^\pm-n)-1}K_0^\pm(n,n\pm j)z^j,\quad n\in\integ.\footnote{We use the convention that $\sum_{j=n}^mc_j=0$ whenever $n>m$.}
\end{equation}
Consequently, the Jost solutions have the following form:
\begin{equation}\label{fplus}
f^+(n,z)=\frac{1}{A^+(n)}\left(z^n+\sum_{m=n+1}^{2N^+-n-1}K_0^+(n,m)z^m\right)
\end{equation}
and
\begin{equation}\label{fminus}
f^-(n,z)=\frac{1}{A^-(n)}\left(z^{-n}+\sum_{m=2N^--n+1}^{n-1}K_0^-(n,m)z^{-m}\right).
\end{equation}
Set $K_0^\pm(n,m)=0$ for $\pm(n-m)\ge 0$ and $\pm(m+n-2N^\pm)\ge0$.  Since $K_0^\pm(n,\cdot)\in\ell_0(\integ,\real)$ we may define the \emph{transformation operators} for the pair $(\tau,\tau_0)$, $\mathcal K^\pm_0:\ell(\integ)\to\ell(\integ)$, by
\begin{align}
\label{K0plusop}(\mathcal K_0^+f)(n)&=\frac{1}{A^+(n)}\left(f(n)+\sum_{m=n+1}^{\infty}K_0^+(n,m)f(m)\right)\\
\label{K0minusop}(\mathcal K_0^-f)(n)&=\frac{1}{A^-(n)}\left(f(n)+\sum_{m=-\infty}^{n-1}K_0^-(n,m)f(m)\right).
\end{align}

Transformation operators are discussed in more detail in section~\ref{sec:transop}.  However, a couple of (standard) facts will be needed in this section, so we list them here.  The kernels $K_0^\pm(n,m)$ satisfy
\begin{equation}
\begin{aligned}
\label{K0pdifeq}K_0^+(n,m+1)-K_0^+(n-1,m)&=b_nK_0^+(n,m) +b_n\delta_{m,n}-K_0^+(n,m-1) \\
 &\qquad+a_n^2K_0^+(n+1,m) +(a_n^2-1)\delta_{m,n+1}
 \end{aligned}
 \end{equation}
\begin{align}
\label{K0psum}K_0^+(n,n+1)=-\sum_{m=n+1}^{N^+}b_m,\\
\label{K0msum}K_0^-(n,n-1)=-\sum_{m=N^-}^{n-1}b_m.
\end{align}
Furthermore, the following equation is true for any $f\in\ell(\integ)$:  \begin{equation}\label{trans0op}
\tau\mathcal K_0^\pm f=\mathcal K_0^\pm\tau_0f.
\end{equation}

We now proceed with defining the central object of scattering theory, the scattering matrix.  First, we define the (modified) Wronskian of $f,g\in\ell(\integ)$ by
\begin{equation*}
[f,g]_n=a_n(f(n)g(n+1)-f(n+1)g(n))
\end{equation*}
which is independent of $n$ whenever $f$ and $g$ solve \eqref{dse} (for the same $z+z^{-1}$) and is nonzero if and only if $f$ and $g$ are linearly independent.  Therefore, we get
\begin{equation*}
[f^\pm(z),f^\pm(z^{-1})]=\pm\frac{1-z^2}{z},\quad z\in\mathbb C\setminus\set{0}
\end{equation*}
so that $\set{f^\pm(z),f^\pm(z^{-1})}$ are linearly independent for $z\neq\pm 1$.  Hence, for every $z\in\mathbb C\setminus\set{0,\pm1}$ we can write 
\begin{equation}\label{fexpand}
f^\pm(n,z)=\alpha(z)f^\mp(n,z^{-1})+\beta^\mp(z) f^\mp(n,z),
\end{equation} where
\begin{equation}\label{alpha}
\alpha(z)=\frac {z[f^+(z),f^-(z)]}{1-z^2},
\end{equation}
and
\begin{equation}\label{beta}
\beta^\pm(z)=\pm\frac{z[f^\mp(z),f^\pm(z^{-1})]}{1-z^2}.
\end{equation}

The scattering matrix is defined by
\begin{equation*}
\mathcal S(z)=
\begin{pmatrix}
\mathcal T(z)& \mathcal{R}^-(z)\\
\mathcal R^+(z)& \mathcal T(z)
\end{pmatrix}
\end{equation*}
where $\mathcal T=\alpha^{-1}$ is the transmission coefficient and $\mathcal R^\pm=\beta^\pm\alpha^{-1}$ are the right and left reflection coefficients, respectively.  Note that $\mathcal T$ and $\mathcal R^\pm$ can be defined at $z=\pm 1$.

We define 
\begin{equation}
\label{w}w(z)=z[f^+(z),f^-(z)]=(1-z^2)\alpha(z)
\end{equation}
and 
\begin{equation}
\label{s}s^\pm(z)=z[f^-(z^{\pm1}),f^+(z^{\mp1})]=(1-z^2)\beta^\pm(z).
\end{equation}  The following lemma contains essential properties of these functions.

\begin{lem}\label{lem:wands}
The functions $w$ and $z\mapsto z^{\pm2N^\pm}s^\pm(z)$ are polynomials with real coefficients whose degrees are at most $\max\set{2,2(N^+-N^-)}$ and $2(N^+-N^-)+1$, respectively.  They have the following properties.
\begin{itemize}
\item[(i)] $w(z)w(z^{-1})+(z-z^{-1})^2=s^\pm(z)s^\pm(z^{-1})$;
\item[(ii)] $\overline{w(z)}=w(\overline z)$, $\overline{s^\pm(z)}=s^\pm(\overline z)$, $s^+(z)=z^2s^-(z^{-1})$;
\item[(iii)] $w(\pm1)=-s^+(\pm1)=-s^-(\pm1)$;
\item[(iv)] $w(0)=A^{-1}$ and $\lim_{z\to0}z^{\pm2N^\pm-1}s^\pm(z)=b_{N^\pm}/A$.
\end{itemize}
\begin{proof}
We will prove first that the given functions are polynomials.  Since the Wronskians defining $w$ and $s^-$ are independent of $n$, we evaluate at $n=N^--1$.  Using \eqref{w}, \eqref{s}, \eqref{g}, and \eqref{K0psum} produces
\begin{equation*}
\begin{aligned}
w(z)&=A^{-1}(g^+(N^--1,z)-z^2g^+(N^-,z))\\
&=A^{-1}(1-z^2)\\
&\quad+A^{-1}\sum_{j=1}^{2(N^+-N^-)+1}[K_0^+(N^--1,N^--1+j)-K_0^+(N^-,N^-+j-2)]z^j,
\end{aligned}
\end{equation*}
and
\begin{equation*}
\begin{aligned}
s^-(z)&=A^{-1}z^{2N^-}(g^+(N^-,z)-g^+(N^--1,z))\\
&=z^{2N^-+1}A^{-1}b_{N^-}\\
&\quad+z^{2N^-+1}A^{-1}\sum_{j=1}^{2(N^+-N^-)}[K_0^+(N^-,N^-+j+1)-K_0^+(N^--1,N^-+j)]z^j,
\end{aligned}
\end{equation*}
where we have also used $K_0^+(\mu,\nu)=0$ for $\mu+\nu\ge2N^+$.  Therefore, $w$ and $z\mapsto z^{-2N^-}s^-(z)$ are polynomials with real coefficients and (iv) is true for them.  Also, the degree of $z\mapsto z^{-2N^-}s^-(z)$ is at most $2(N^+-N^-)+1$.  The corresponding claims concerning $s^+$ can be obtained by evaluating the Wronskian at $n=N^+$ and arguing in a similar fashion.

We see that $w$ has degree at most $\max\set{2,2(N^+-N^-)+1}$.  However, we claim the coefficient corresponding to $2(N^+-N^-)+1$ is zero if $N^+\neq N^-$, reducing the degree by 1.  This coefficient is $$A^{-1}(K_0^+(N^--1,2N^+-N^-)-K_0(N^-,2N^+-N^--1))$$ and can be calculated in the following way.     Evaluate \eqref{K0pdifeq} at $n=N^-$ and $m=2N^+-N^-$ while keeping in mind that $K_0^+(\nu,\mu)=0$ for $\mu+\nu\ge2N^+$ and $a_\nu-1=b_\nu=0$ for $\nu\le N^-$.  The result is $$K_0^+(N^--1,2N^+-N^-)-K_0(N^-,2N^+-N^--1)=b_{N^-}\delta_{2N^+-N^-,N^-}.$$  Unless $N^+=N^-$, the right hand side is 0.  When $N^+=N^-$, we find $w(z)=1-z^2-b_{N^-}z$ since $A=1$.

Part (i) is proven using the Pl\"ucker identity \cite[2.169]{Teschl2000}.  The first two statements of part (ii) are true of every polynomial with real coefficients.  Part (iii) and the third statement of part (ii) follow from \eqref{w} and \eqref{s}.
\end{proof}
\end{lem}

The zeros of $w$ are important.  For if $w(z_0)=0$ and $\abs{z_0}<1$, then $f^+(\cdot,z_0)\in\ell^2(\integ)$ being proportional to $f^-(\cdot,z_0)$.  Therefore, $\lambda(z_0)=z_0+z_0^{-1}$ is an eigenvalue of the operator $J$.  Since we are only concerned with zeros of $w$ and not their transformation under $\lambda(z)$, we will refer to those zeros inside the unit disk as eigenvalues.  In addition, the zeros of $w$ outside the unit disk are called \emph{resonances}.  Combining parts (i) and (ii) of the above lemma shows that $w$ cannot vanish on the unit circle except possibly at $\pm1$.  If $w(1)=0$ or $w(-1)=0$, we say there is a \emph{half-bound state} at $z=1$ or $z=-1$, respectively.  Observe that part (i) in the above lemma implies that $w$ and $s^\pm$ cannot vanish simultaneously except at $z=1$ or $z=-1$.  Hence, the poles of $\mathcal T$ and $\mathcal R^\pm$ are precisely the zeros of $w$ in $\mathbb C\setminus\set{\pm1}$.  In addition, the zeros of $s^\pm$ are the zeros of $\mathcal R^\pm$ in the same set.

Let $w_j$ be the zeros of $w$ and $s_j$ the non-zero zeros of $s^-$.  Then, the zeros of $s^+$ are $s_j^{-1}$ by Lemma~\ref{lem:wands} (ii).  By part (iv) of the same lemma, we conclude
\begin{align}
\label{wfact}w(z)&=\frac1A\prod_{j=1}^{M_w}\left(1-\frac{z}{w_j}\right),\quad M_w\le\max\set{2,2(N^+-N^-)}\\
\label{sfact}s^\pm(z)&=\frac{b_{N^\pm}}{A}z^{1\mp2N^{\pm}}\prod_{j=1}^{M_s}\left(1-\frac{z}{s_j^{\mp1}}\right),\quad M_s\le2(N^+-N^-).
\end{align}
We will need these factorizations throughout the rest of the paper and, in particular, for the proof of Theorem~\ref{thm:unique}.

Turning to the inverse problem, we recall that the \emph{norming constants}, $\gamma^\pm$, are defined by
\begin{equation}\label{normconst}
(\gamma^\pm)^{-1}=\sum_{n\in\integ}\abs{f^\pm(n,z_j)}^2,
\end{equation}
where $z_j$ is an eigenvalue.  Furthermore, the following identity is true
\begin{equation}\label{resT}
\text{res}(\mathcal T,z_j)=-z_j\gamma^\pm_j\mu_j^{\mp1},
\end{equation} 
where $\mu_j$ is defined by $f^-(\cdot,z_j)=\mu_jf^+(\cdot,z_j)$.

The standard inverse scattering result states that $\set{\mathcal R^+, z_j, \gamma_j^+}$ or $\set{\mathcal R^-,\, z_j,\, \gamma_j^-}$ determine the sequences $a$ and $b$ uniquely.  We note that the reflection coefficients, $\mathcal R^\pm$, need to be known only on the unit circle.  Indeed, they can only be defined there under the usual scattering conditions on $a$ and $b$.  Since we assume $a-1$ and $b$ are finitely supported, we find that only the reflection coefficients (defined on $\mathbb C$) are needed.
\begin{lem}\label{lem:refcoeffonly}
If the coefficients $a$ and $b$ of a JDE satisfy $a-1,\, b\in\ell_0(\integ,\real)$, then they are uniquely determined by the left or right reflection coefficient.
\begin{proof}
For $a-1,\, b\in\ell_0(\integ,\real)$ we know that $\mathcal R^\pm(z)=\beta^\pm(z)\alpha^{-1}(z)$ is a meromorphic function in $\mathbb C$ whose poles in the unit disk are the eigenvalues, $z_j$.  Moreover, since these poles are simple \cite{Teschl2000} and $\beta^\pm(z_j)=\mu_j^{\pm1}$, we find 
\begin{equation}\label{resR}
\text{res}(\mathcal R^\pm,z_j)=\lim_{z\to z_j}\beta^\pm(z)\mathcal T(z)(z-z_j)=\beta^\pm(z_j)\text{res}(\mathcal T,z_j)=-z_j\gamma^\pm_j,
\end{equation}
by using \eqref{resT}.
Hence, the reflection coefficients determine the sets $\set{\mathcal R^\pm,\, z_j,\, \gamma_j^\pm}$ and, therefore, the sequences $a$ and $b$.
\end{proof}
\end{lem}

\begin{remark}
A more general form of this lemma can be proved assuming only that $a_n-1$ and $b_n$ vanish either for $n>N^+$ or $n<N^-$, i.e. they are supported on a half-lattice.  Then, $\mathcal R^+$ (resp. $\mathcal R^-$) may be defined meromorphically in the unit disk.  The proof mirrors the one above exactly.  See \cite{Markushevich1985} for the continuous analogue.
\end{remark}

The final ingredient for the proof of Theorem~\ref{thm:unique} is to determine the effect of a translation of $a$ and $b$ on the reflection coefficient.

\begin{lem}\label{lem:translate}
Suppose $\tau$ and $\tilde\tau$ satisfy Assumption~\ref{assume:finitesupp}.  Then, there exists $j_0\in\integ$ such that $\tilde a_n=a_{n-j_0}$ and $\tilde b_n=b_{n-j_0}$ if and only if $\tilde{\mathcal R}^\pm(z)=z^{\mp2j_0}\mathcal R^\pm(z)$.
\begin{proof}
Let $N^\pm$ and $\tilde N^\pm$ be the integers given by Assumption~\ref{assume:finitesupp} for $\tau$ and $\tilde\tau$, respectively.  Suppose there is a $j_0\in\integ$ such that $\tilde a_n=a_{n-j_0}$ and $\tilde b_n=b_{n-j_0}$.  Thus, $\tilde N^\pm=N^\pm+j_0$.  Let $f^\pm(n,z)$ and $\tilde f^\pm(n,z)$ be the Jost solutions for $\tau$ and $\tilde\tau$, respectively.  Upon inspection, it is clear that $\tilde f^\pm(n,z)=z^{\pm j_0}f^\pm(n-j_0,z)$.  By substituting these into \eqref{w} and \eqref{s}, we find that $\tilde w(z)=w(z)$ and $\tilde s^\pm(z)=z^{\mp 2j_0}s^\pm(z)$.  Therefore, the reflection coefficients have the desired form.

On the other hand, if $\tilde{\mathcal R}^\pm(z)=z^{\mp2j_0}\mathcal R^\pm(z)$, then Lemma~\ref{lem:refcoeffonly} guarantees that $\tilde a_n=a_{n-j_0}$ and $\tilde b_n=b_{n-j_0}$.
\end{proof}
\end{lem}

\begin{thm}\label{thm:unique}
The sequences $a$ and $b$ with $a-1,b\in\ell_0(\integ,\real)$ of a JDE are determined uniquely up to a translation by the zeros and poles of one of the reflection coefficients, $\mathcal R^\pm$, whenever there is at most one half-bound state or at least one eigenvalue.
\begin{proof}
We prove the case for $\mathcal R^-$ as the one for $\mathcal R^+$ is similar.  Due to Lemmas \ref{lem:refcoeffonly} and \ref{lem:translate}, we only need to show that the zeros and poles of $\mathcal R^-$ are sufficient to determine it up to its order at zero.  By the factorizations \eqref{wfact} and \eqref{sfact}, the reflection coefficient is determined by its zeros and poles up to the factor $b_{N^-}z^{1+2N^-}$.  We, therefore, need only determine $b_{N^-}$.  Suppose there is at most one half-bound state, and, without loss of generality, that if there is one it is at $z=-1$.  Lemma~\ref{lem:wands} implies $\mathcal R^-(1)=-1$ in this case.  By \eqref{wfact} and \eqref{sfact}, $b_{N^-}$ can be determined.  However, if there are two half-bound states we cannot use this method because $w(\pm1)=0=-s^-(\pm 1)$.  Therefore, we find $b_{N^-}^2$ by evaluating the equation in the Lemma~\ref{lem:wands} (i) at an eigenvalue, $w_j$.  By \eqref{resR}, we have
\begin{equation*}
-w_j^{-1}\text{res}(R^-,w_j)=\gamma^-_j>0,
\end{equation*}
Using \eqref{wfact} and \eqref{sfact}, this inequality means the sign of $b_{N^-}$ is determined.
\end{proof}
\end{thm}
We note that Korotyaev, in \cite{Korotyaev2011} Theorem 1.3.i, has a result in much the same vein as Theorem~\ref{thm:unique}.  First, he excludes the possibility of a shift by prescribing \textit{a priori} the endpoints of the supports of $a-1$ and $b$ (see equation $(1.1)$ in the referenced article).  Next, the data given for the inversion are the polynomial $w$ and a finite sequence, $\sigma$, whose values are taken from the set $\set{-1,1}$ (subject to some characterization constraints).  By Lemma~\ref{lem:wands} (i) $s^-(z)s^-(1/z)$ is known.  Therefore we know the zeros of $s$ up to an exponent of $\pm 1$, i.e. if $\zeta$ is a zero of $s^-(z)s^-(1/z)$, then $\zeta$ or $\zeta^{-1}$ is a zero of $s^-$.  The sequence $\sigma$ is used to choose this exponent and to determine the sign of $b_{N^-}$.  Therefore, the reflection coefficient, $\mathcal R^-$, is determined.  Knowing the sequence $\sigma$ is the same as knowing the zeros of $s^-$, so, in this sense, our result is not new and we do not claim originality.  Our goal is a stability result and the zeros of $s^-$ are easier to work with than $\sigma$ in this context.

\section{Stability}\label{sec:stability}
For simplicity (and without loss of generality) we set $N^-=0$ and $N^+=N\in\mathbb N_0$.  Furthermore, we require that the coefficients of a JDE lie within a ball of radius $Q\ge1$ in $\ell^\infty(\integ)$.  Hence, we make the following definition.
\begin{definition}\label{B0}
Let $N\in\mathbb N_0$ and $Q\ge 1$ be given.  The set $B_0(N,Q)$ consists of all JDE, $\tau$, that satisfy Assumption~\ref{assume:finitesupp} with $N^-=0$, $N^+=N$, and for which the sequences $a$ and $b$ defining $\tau$ obey $\norm{a}_\infty,\, \norm{1/a}_\infty,\, \norm{b}_\infty\le Q$.

In addition, given $\delta\in(0,1)$ we say that $\tau\in B_\delta(N,Q)$ if $\tau\in B_0(N,Q)$ and $\abs{b_0},\abs{s(1)}\ge\delta$.
\end{definition}

\begin{remark}For $\tau\in B_\delta(N,Q)$, the assumption $\abs{b_0}\ge\delta$ ensures that there is no $\tilde\tau\in B_\delta(N,Q)$ for which $\tilde b$ is a translation of $b$.  In addition, the assumption $\abs{s(1)}\ge\delta$ means that there is no half-bound state at $z=1$.  According to Theorem~\ref{thm:unique}, $\tau\in B_\delta(N,Q)$ is uniquely determined by the zeros and poles of its left reflection coefficient.

We also note that it is no restriction to assume $\delta\in(0,1)$ since $B_\delta(N,Q)\supset B_{\delta'}(N,Q)$ whenever $\delta\le\delta'$. 
\end{remark}

We will deal exclusively with the transformation operator $\mathcal K_0^+$ and its kernel $K_0^+(n,m)$ as opposed to $\mathcal K_0^-$, and with the reflection coefficient $\mathcal R^-$ and the associated function $s^-$, as opposed to $\mathcal R^+$ and $s^+$.  For this reason, we will drop the superscripts for these objects in the sequel, i.e. $\mathcal K_0=\mathcal K_0^+$, $K_0=K_0^+$, $\mathcal R=\mathcal R^-$, and $s=s^-$. 
\subsection{More on transformation operators}\label{sec:transop}
In this subsection, we follow \cite[Sec. 3.1]{MNSW10} closely and will usually refer the reader there for proofs.  We begin with noting that for $\tau\in B_0(N,Q)$ the inverse $\mathcal L_0=\mathcal K_0^{-1}$ exists and has a kernel $L_0(n,m)$ such that
\begin{equation}\label{L0}
(\mathcal L_0f)(n)=A^+(n)\left(f(n)+\sum_{m=n+1}^\infty L_0(n,m)f(m)\right),\quad f\in\ell(\integ),
\end{equation}
and $L_0(n,m)=0$ for $m\le n$ and $m+n\ge2N$.  We will call $\mathcal L_0$ a transformation operator for the pair $(\tau_0,\tau)$.

Let $\tau,\tilde\tau\in B_0(N,Q)$ with the transformation operators $\mathcal K_0$, $\mathcal L_0$, $\tilde{\mathcal K}_0$, and $\tilde{\mathcal L}_0$.  The transformation operator for the pair $(\tau,\tilde\tau)$ is $\mathcal K=\tilde{\mathcal K}_0\mathcal L_0$.  Applying \eqref{K0plusop} and \eqref{L0}, we find that 
\begin{equation}\label{Kop}
(\mathcal Kf)(n)=\frac{A^+(n)}{\tilde A^+(n)}\left(f(n)+\sum_{m=n+1}^\infty K(n,m)f(m)\right),
\end{equation}
where the kernel, $K(n,m)$, satisfies
\begin{equation}\label{K}
\begin{aligned}
K(n,m)&=L_0(n,m)+\frac{A^+(m)}{A^+(n)}\tilde K_0(n,m)\\
&\qquad +\sum_{k=n+1}^{m-1}\frac{A^+(k)}{A^+(n)}\tilde K_0(n,k)L_0(k,m),\quad m>n.
\end{aligned}
\end{equation}
Thus $K(n,m)=0$ for $m\le n$ and $m+n\ge2N$.  Note that the kernel of the transformation operator in \cite{MNSW10}, $K_1(n,m)$, is slightly different than ours; they are related via $K_1(n,m)=A^+(n)(\tilde A^{+}(n))^{-1}K(n,m)$.  

The connection between $K$, $\tilde K_0$, $K_0$, and $L_0$ is given by
\begin{equation}\label{K0diff}
\begin{aligned}
K(n,m)=\frac{A^+(m)}{A^+(n)}&\left[\tilde K_0(n,m)-K_0(n,m)\right]\\ &+\sum_{k=n+1}^{m-1}\frac{A^+(k)}{A^+(n)}L_0(k,m)\left[\tilde K_0(n,k)-K_0(n,k)\right].
\end{aligned}
\end{equation}

By \eqref{trans0op}, we deduce 
\begin{equation}\label{transop}
\tilde\tau\mathcal Kf=\mathcal K\tau f,\quad f\in\ell(\integ).
\end{equation}
Define
\begin{equation*}
(\square K)(n,m)=a_{n-1}K(n-1,m)+\frac{\tilde a_n^2}{a_n}K(n+1,m)-a_mK(n,m+1)-a_{m-1}K(n,m-1).
\end{equation*}
Substituting \eqref{Kop} into \eqref{transop} implies (after a direct calculation)
\begin{equation}\label{Kdiffeq}
(\square K)(n,m)=(b_m-\tilde b_n)K(n,m)+\frac{1}{a_n}(a_n^2-\tilde a_n^2)\delta_{m,n+1}+(b_n-\tilde b_n)\delta_{m,n}.
\end{equation}
Furthermore, setting $q_{n,j}=\sum_{k=n+1}^j(b_{k+1}-\tilde b_k)$, we find
\begin{equation}
\label{K1}K(n,n+1)=\frac{1}{a_n}\sum_{j=n+1}^N(b_j-\tilde b_j),\quad n\le N-1
\end{equation}
and
\begin{equation}
\label{K2}K(n,n+2)=\frac{1}{a_na_{n+1}}\sum_{j=n+1}^{N-1}(q_{n,j}(b_{j+1}-\tilde b_{j+1})+a_j^2-\tilde a_j^2),\quad n\le N-2.
\end{equation}

By Lemmas 3.1 and 3.3 in \cite{MNSW10}, there exist two constants, $C_0$ and $C_1$, each depending only on $N$ and $Q$ such that for $n\ge-1$, $n<m<2N-n$, the following inequalities are valid:
\begin{align}
\label{crude}\abs{K(n,m)}&\le C_0,\\
\label{invest}\abs{K(n,m)}&\le C_1\max_{2\le j\le 2N}\abs{K(-1,j)}.
\end{align}
We refer the reader to the cited paper for the details.

The inequality \eqref{invest} will be integral in the proof of Theorem~\ref{thm:stability}.  In particular, since \eqref{crude} applies to $L_0$ (take $\tilde\tau=\tau_0$), we have by \eqref{K0diff} 
\begin{equation*}
\abs{K(-1,m)}\le C\sum_{j=0}^m\abs{K_0(-1,m)-\tilde K_0(-1,m)}
\end{equation*}
for some $C=C(N,Q)$.  Hence, inequality \eqref{invest} implies there is a $C$ depending only on $N$ and $Q$ such that
\begin{equation}\label{Kest}
\abs{K(n,m)}\le C\sum_{j=0}^{2N}\abs{K_0(-1,m)-\tilde K_0(-1,m)}.
\end{equation}

\subsection{Estimates for $w-\tilde w$ and $s-\tilde s$}\label{sec:wands}
We begin with a couple propositions that will be used in this section.
\begin{prop}\label{prop1}
Let $Z>0$ and $M\in\mathbb N$.  If $z_j,\tilde z_j\in\mathbb C$ and $\abs{z_j},\abs{\tilde z_j}\le Z$ for $1\le j\le M$, then
\begin{equation*}
\abs{\prod_{j=1}^M(1-z_j)-\prod_{j=1}^M(1-\tilde z_j)}\le(1+Z)^{M-1}\sum_{j=1}^{M}\abs{z_j-\tilde z_j}.
\end{equation*}
\begin{proof}
The statement is clearly true for $M=1$.  Assuming it is true for $M$, we estimate
\begin{align*}
\abs{\prod_{j=1}^{M+1}(1-z_j)-\prod_{j=1}^{M+1}(1-\tilde z_j)}&\le\abs{1-z_{M+1}}\abs{\prod_{j=1}^{M}(1-z_j)-\prod_{j=1}^M(1-\tilde z_j)}\\
&\qquad\qquad+\abs{z_{M+1}-\tilde z_{M+1}}\abs{\prod_{j=1}^M(1-\tilde z_j)}\\
&\le(1+Z)^M\sum_{j=1}^{M}\abs{z_j-\tilde z_j}+(1+Z)^M\abs{z_{M+1}-\tilde z_{M+1}}
\end{align*}
proving the claim.
\end{proof}
\end{prop}

\begin{prop}\label{prop2}
Let $P(z)=\sum_{j=0}^Mc_jz^j$ where $\abs{c_j}\le C$, $c_0\neq 0$, and $M\in\mathbb N$.  If $p_0$ is a zero of $P$, then $\abs{c_0}/(MC)\le p_0$.
\begin{proof}
For $\abs{z}\le 1$, we have $\abs{P(z)-c_0}\le MC\abs{z}$.  Thus, $\abs{P(z)}\ge\abs{c_0}-MC\abs{z}$.  Therefore, $P(z)$ cannot vanish for $\abs{z}<\abs{c_0}/(MC)$.
\end{proof}
\end{prop}

Let $\tau,\tilde\tau\in B_0(N,Q)$ for $N\in\mathbb N_0$ and $Q>0$.  The JDE $\tau$ has associated with it the Jost solutions $f^\pm$, the kernel $K_0(n,m)$, the number $A$ (see \eqref{A}), and the polynomials $w$ and $s$ while $\tilde\tau$ has its corresponding objects marked with a tilde.  The kernel of the transformation operator associated with $(\tau,\tilde\tau)$ is $K(n,m)$.  

Let $w_j$, $j=1\dots M_w$ and $\tilde w_j$, $j=1\dots M_{\tilde w}$ be the zeros of $w$, respectively $\tilde w$, listed according to multiplicity and by increasing modulus where $M_w,M_{\tilde w}\le \max\set{2,2N}$.  We first apply Proposition~\ref{prop2} to $Aw$ and $\tilde A\tilde w$.  In the proof of Lemma~\ref{lem:wands} we found that the coefficients of $Aw$ and $\tilde A\tilde w$ depend only on the values of $K(n,m)$ and $\tilde K(n,m)$ for $n\in\set{-1,0}$, respectively.  Inequality \eqref{crude} implies that, in the notation of Proposition~\ref{prop2}, if $P$ is either $Aw$ or $\tilde A\tilde w$, then $\abs{c_j}\le C_0$.  Furthermore, $c_0=1$ by Lemma~\ref{lem:wands} (iv).  Therefore, there is a positive $\gamma_w=\gamma_w(N,Q)\le 1$ such that
\begin{equation}\label{gammaw}
\gamma_w\le\abs{w_j},\abs{\tilde w_j}.
\end{equation}

\begin{lem}\label{lem:wdiff}
Let $N\in\mathbb N_0$ and $Q\ge1$.  Then there is a positive constant $C=C(N,Q)$ such that for any $R\ge 1$ and any $\eps>0$ the following statement is true.  If $\tau,\tilde\tau\in B_0(N,Q)$ have corresponding functions $w$ and $\tilde w$ with the same number, $m_w$, of zeros in the disk $\abs{z}<R$ and $\abs{w_j-\tilde w_j}\le\eps$, $j=1,\dots,m_w$,  then the following inequality holds for any $z$ in closed unit disk:
\begin{equation*}
\abs{Aw(z)-\tilde A\tilde w(z)}\le C(\eps +\frac1R).
\end{equation*}
\begin{proof}
By \eqref{wfact}, we find
\begin{align*}
\abs{Aw(z)-\tilde A\tilde w(z)}&
\le\prod_{j=1}^{m_w}\abs{1-\frac{z}{\tilde w_j}}\left[\abs{\prod_{j=m_w+1}^{M_w}\left(1-\frac{z}{w_j}\right)-1}+\abs{\prod_{j=m_w+1}^{M_{\tilde w}}\left(1-\frac{z}{\tilde w_j}\right)-1}\right]\\
&\qquad+\prod_{j=m_w+1}^{M_w}\abs{1-\frac{z}{w_j}}\abs{\prod_{j=1}^{m_w}\left(1-\frac{z}{w_j}\right)-\prod_{j=1}^{m_w}\left(1-\frac{z}{\tilde w_j}\right)}.
\end{align*}
We apply Proposition~\ref{prop1}  and \eqref{gammaw} to find
\begin{align*}
\abs{Aw(z)-\tilde A\tilde w(z)}&\le (1+\gamma_w^{-1})^{M_w-1}\frac{m_w}{\gamma_w^2}\eps+(M_w-m_w)(1+\gamma_w^{-1})^{M_w-1}\frac1R\\
&\qquad+(M_{\tilde w}-m_w)(1+\gamma_w^{-1})^{M_{\tilde w}-1}\frac1R.
\end{align*}
Since $\gamma_w$, $M_w$, and $M_{\tilde w}$ depend only on $N$ and $Q$, the lemma is proven.
\end{proof}
\end{lem}

We now turn to $s$ and $\tilde s$.  We will henceforth fix $\delta\in(0,1)$ and take $\tau,\tilde\tau\in B_\delta(N,Q)$.  As we did in the case of $w$ and $\tilde w$, we let $s_j$, $j=1\dots M_s$ and $\tilde s_j$, $j=1\dots M_{\tilde s}$ be the nonzero zeros of $s$, respectively $\tilde s$, listed according to multiplicity and by increasing modulus where $M_s,M_{\tilde s}\le \max\set{2,2N}$.  We again apply Proposition~\ref{prop2} to $As$ and $\tilde A\tilde s$.  Again Lemma~\ref{lem:wands} implies that the coefficients of $As$ and $\tilde A\tilde s$ also depend only on the values of $K(n,m)$ and $\tilde K(n,m)$ for $n\in\set{-1,0}$, respectively.  Therefore,  inequality \eqref{crude} implies that each coefficient is bounded by $C_0$.  However,  $c_0$ is either $b_0$ or $\tilde b_0$ by Lemma~\ref{lem:wands} (iv).  Therefore, there is a positive $\gamma_s=\gamma_w(N,Q,\delta)\le 1$ such that
\begin{equation}\label{gammas}
\gamma_s\le\abs{s_j},\abs{\tilde s_j}.
\end{equation}

\begin{lem}\label{lem:sdiff}
Let $N\in\mathbb N_0$, $Q\ge1$, and $\delta\in(0,1)$.  Then there is a positive constant $C=C(N,Q,\delta)$ such that for any $R\ge 1$ and any $\eps>0$ the following statement is true.  If $\tau,\tilde\tau\in B_0(N,Q)$ have corresponding functions $s$ and $\tilde s$ with the same number, $m_s$, of zeros in the disk $\abs{z}<R$ and $\abs{s_j-\tilde s_j}\le\eps$, $j=1,\dots,m_s$,  then the following inequality holds for any $z$ in closed unit disk:
\begin{equation*}
\abs{As(z)-\tilde A\tilde s(z)}\le C(\eps +\frac1R).
\end{equation*}

\begin{proof}
The proof proceeds similarly to that of Lemma~\ref{lem:wdiff}.  By \eqref{sfact}, we have
\begin{align*}
\abs{As(z)-\tilde A\tilde s(z)}&\le\abs{b_0z}\prod_{j=m_s+1}^{M_s}\abs{1-\frac{z}{s_j}}\abs{\prod_{j=1}^{m_s}\left(1-\frac{z}{s_j}\right)-\prod_{j=1}^{m_s}\left(1-\frac{z}{\tilde s_j}\right)}\\
&\qquad+\abs{b_0z}\prod_{j=1}^{m_s}\abs{1-\frac{z}{\tilde s_j}}\abs{\prod_{j=m_s+1}^{M_s}\left(1-\frac{z}{s_j}\right)-\prod_{j=m_s+1}^{M_{\tilde s}}\left(1-\frac{z}{\tilde s_j}\right)}\\
&\qquad\qquad+\abs{z}\prod_{j=m_s+1}^{M_{\tilde s}}\abs{1-\frac{z}{\tilde s_j}}\abs{b_0-\tilde b_0}.
\end{align*}
The first two terms in the sum are bounded by $C(\eps +1/R)$ where $C=C(N,Q,\delta)$.  The $\delta$ dependence comes from estimating $\abs{s_j^{-1}}$ and $\abs{\tilde s_j}^{-1}$ by $\gamma_s^{-1}$.  These bounds are proven similarly to the proof of Lemma~\ref{lem:wdiff}.

In order to estimate $\abs{b_0-\tilde b_0}$, we write 
\begin{align*}
b_0-\tilde b_0&= \frac{As(1)}{\prod_{j=1}^{M_s}\left(1-\frac{1}{ s_j}\right)}-\frac{\tilde A\tilde s(1)}{\prod_{j=1}^{M_{\tilde s}}\left(1-\frac{1}{\tilde s_j}\right)}\\
&=\frac{b_0}{As(1)}\left[\tilde A\tilde w(1)-Aw(1)+\tilde b_0\left(\prod_{j=1}^{M_s}\left(1-\frac{1}{ s_j}\right)-\prod_{j=1}^{M_{\tilde s}}\left(1-\frac{1}{\tilde s_j}\right)\right)\right]
\end{align*}
after using \eqref{sfact} more than once as well as Lemma~\ref{lem:wands} (iii).  The first term is estimated using Lemma~\ref{lem:wdiff}, and the second by employing the method described in the preceding paragraph.  We, therefore, find that $\abs{b_0-\tilde b_0}\le C(\eps+1/R)$ for some constant $C$ depending on $N$, $Q$, and $\delta$.  This completes the proof.
\end{proof}
\end{lem}

\begin{remark}
From Definition~\ref{B0}, we know that $\gamma_s$ is proportional to $\delta$.  Therefore, the constant $C$ in the previous lemma is proportional to $\delta^{-k}$ for some $k\in\mathbb N$ depending on $N$.
\end{remark}

\subsection{The main result}\label{sec:main}
We are now ready to prove the following.
\begin{thm}\label{thm:stability}
Let $N\in\mathbb N_0$, $Q\ge1$, and $\delta\in(0,1)$ be given.  Then there is a positive constant $C=C(N,Q,\delta)$ such that for any $R\ge1$ and any $\eps>0$ the following statement is true.  If $\tau,\tilde\tau\in B_\delta(N,Q)$ have reflection coefficients whose zeros and poles in the disk $\abs{z}<R$ are, respectively, $\eps$-close, then
\begin{align*}
\abs{b_n-\tilde b_n}&\le C\left(\eps+\frac1R\right)\\
\abs{a_n^2-\tilde a_n^2}&\le C\left(\eps+\frac1R\right).
\end{align*}

\begin{proof}
By \eqref{fexpand}, \eqref{fplus}, \eqref{fminus}, \eqref{w}, and \eqref{s} we have
\begin{align}
\label{fatminus1}Af^+(-1,z)-\tilde A\tilde f^+(-1,z)&=\frac{z^{-1}(Aw(z)-\tilde A\tilde w(z))+z(As(z)-\tilde A\tilde s(z))}{1-z^2}\\
\label{fatminus2}&=\sum_{m=0}^{2N}(K_0(-1,m)-\tilde K_0(-1,m))z^m.
\end{align}
So, Cauchy's theorem and \eqref{fatminus2} imply that for $0\le m\le 2N$ and any $r>0$,
\begin{equation}\label{KminustildeK}
K_0(-1,m)-\tilde K_0(-1,m)=\frac{1}{2\pi i}\int_{\abs{z}=r}\frac{Af^+(-1,z)-\tilde A\tilde f^+(-1,z)}{z^{m+1}}\, dz.
\end{equation}
Choosing $r=1/2$, using \eqref{fatminus1}, and applying Lemmas \ref{lem:wdiff} and \ref{lem:sdiff} yields
\begin{equation*}
\abs{K_0(-1,m)-\tilde K_0(-1,m)}\le C(\eps +\frac1R)
\end{equation*}
for some $C=C(N,Q,\delta)$.  

Now we apply \eqref{Kest} to find 
\begin{equation}\label{Kest1}
\abs{K(n,m)}\le C(\eps +1/R).
\end{equation}
Starting with $n=N-1$ and decreasing by one until $n=-1$ in \eqref{K1}, we use \eqref{Kest1} to establish the desired estimate of $\abs{b_n-\tilde b_n}$.  Having this estimate, we use it and \eqref{Kest1} to obtain the bound on $\abs{a_n^2-\tilde a_n^2}$ from \eqref{K2} starting at $n=N-2$ and proceeding one at a time to $n=-1$.  
\end{proof}
\end{thm}

\noindent\textbf{Acknowledgement}  The author gratefully thanks Rudi Weikard for insights and critiques given during the preparation of this work.

\bibliographystyle{plain}
\bibliography{discrete,myref,stability}

\end{document}